\documentclass[useAMS,usenatbib]{mn2e}
\twocolumn
\usepackage{bm}
\usepackage{amsfonts}
\usepackage{amsmath}
\usepackage{graphicx}
\usepackage{subfigure}
\usepackage{times}

\def\gsim{\;\lower4pt\hbox{${\buildrel\displaystyle >\over\sim}$}\;}
\def\lsim{\;\lower4pt\hbox{${\buildrel\displaystyle <\over\sim}$}\;}
\def\grls{\;\lower4pt\hbox{${\buildrel\displaystyle >\over <}$}\;}
\title[Dark Matter in Satellite Dwarf Galaxy Segue 1]
{Dark matter dominated dwarf disc galaxy Segue 1}
\author[M. Xiang-Gruess, Y.-Q. Lou, W. J. Duschl]
  {Meng Xiang-Gruess$^{1}$, Yu-Qing Lou$^{2,\ 3}$, Wolfgang J. Duschl$^{1,\ 4}$\\
$^1$ Christian-Albrechts-Universit\"at zu Kiel, Institut f\"ur Theoretische Physik
     und Astrophysik, Leibnizstr. 15, 24118 Kiel, Germany\\
$^2$ Physics Department and Tsinghua Centre for Astrophysics
(THCA), Tsinghua University, Beijing, 100084, China\\
$^3$ National Astronomical Observatories, Chinese Academy of Sciences,
     A20, Datun Road, Beijing 100012, China\\
$^4$ Steward Observatory, The University of Arizona, 933 North Cherry Ave,
     Tucson, AZ 85721, USA}
\date{Accepted 2009 September 16.
    Received 2009 September 6; in original form 2009 August 19.}
\pagerange{\pageref{firstpage}--\pageref{lastpage}} \pubyear{2009}

\def\LaTeX{L\kern-.36em\raise.3ex\hbox{a}\kern-.15em
    T\kern-.1667em\lower.7ex\hbox{E}\kern-.125emX}

\begin{document}
\label{firstpage}
\maketitle

\begin{abstract}
Several observations reveal that dwarf galaxy Segue 1 has a dark
matter (DM) halo at least $\sim 200$ times more massive than its
visible baryon mass of only $\sim 10^3\ \rmn{M_\odot}$. The baryon
mass is dominated by stars with perhaps an interstellar gas mass
of $\lsim 13\ \rmn{M_\odot}$. Regarding Segue 1 as a {\it dwarf
disc galaxy} by its morphological appearance of long stretch, we
invoke the dynamic model of Xiang-Gruess, Lou \& Duschl (XLD) to
estimate its physical parameters for possible equilibria with and
without an isopedically magnetized gas disc. We estimate the range
of DM mass and compare it with available observational inferences.
Due to the relatively high stellar velocity dispersion compared to
the stellar surface mass density, we find that a massive DM halo
would be necessary to sustain disc equilibria. The required DM
halo mass agrees grossly with observational inferences so far. For
an isopedic magnetic field in a gas disc, the ratio $f$ between
the DM and baryon potentials depends strongly on the magnetic
field strength.
%leading to the conclusion that a very massive
%DM halo may have two possible explanations.
Therefore, a massive DM halo is needed to counteract either the
strong stellar velocity dispersion and rotation of the stellar
disc or the magnetic Lorentz force in the gas disc. By the radial
force balances, the DM halo mass increases for faster
%with the increase of the
disc rotation.
\end{abstract}


\section{Introduction}

%The dwarf galaxy
Segue 1 has been scrutinized among other Milky Way (MW) satellites
(e.g. Putman et al. 2008; Martin et al. 2008; Geha et al. 2009)
since its recent discovery a few years ago (Belokurov et al.
2007). There is an ongoing debate on the classification of Segue
1, i.e. whether Segue 1 is a dwarf galaxy or a stellar cluster.
Belokurov et al. (2007) and Niederste-Ostholt et al. (2009) argue
that Segue 1 is likely a stellar cluster with a distorted outer
stellar part whereas Geha et al. (2009) suggest that Segue 1 is a
dwarf galaxy.
The most remarkable result of Geha et al. (2009) is the inferred
dark matter (DM) halo of Segue 1 which is up to a factor $f\sim
2000$ times its baryon mass. The lower limit for this ratio $f$
between DM and visible baryon matter is $\sim 200$; this is
extraordinary as such ratio between DM and baryon mass in normal
disc galaxies is typically $f\sim 10$. In spite of observational
uncertainties, we may presume that Segue 1 has an unusually large
ratio $f$ calling for further observational and theoretical
confirmations. In this Letter, we presume Segue 1 as a {\it dwarf
disc galaxy}, apply our composite model for disc galaxies
(Xiang-Gruess, Lou \& Duschl 2009; XLD hereafter) and focus on
relevant aspects of this DM issue as well as the possible role of
a magnetic field in its interstellar gas disc.

%{\bf Cosmological implications for the presence of
%Segue 1 type of systems in the Universe.} \newline

Ultra-faint dwarf galaxies like Segue 1, which have such large
fractions or amounts of DM, bear profound implications for the
formation and evolution of galaxies.
In the framework of cold DM cosmology, massive galaxies such as
our MW are predicted to be accompanied by a large number of
DM-dominated satellite halos. Extensive observations in the 1990s
(e.g. Kauffman et al. 1993;
%, White, Guiderdoni 1993;
Willman et al. 2005)
%Klypin et al. 1999; Moore et al. 1999
however have revealed a much smaller number of such satellites.
This discrepancy is known as the \textit{missing satellite
problem} (e.g. Klypin et al. 1999; Moore et al. 1999).

%{\bf
The recent results for Segue 1 and other dwarf satellites of our
MW have several implications. For example, the earlier hypothesis
that all dwarf spheroidals (dSphs) are embedded in DM halos of the
same mass (Mateo et al. 1993) must be treated with care for
ultra-faint dwarf galaxies, as they do not fit into the predicted
curves for the mass-to-light ratio (Simon \& Geha 2007).
%For example, the earlier hypothesis that all dwarf spheroidals
%(dSphs) are embedded in DM halos of the same mass must be
%abandoned since
Here, dwarf galaxies like Segue 1 have DM halos that are much more
massive than the baryon mass. Another example is that, by applying
the results for the ultra-faint dwarf galaxies such as Segue 1,
Simon \& Geha (2007) were able to provide a possible solution to
the missing satellite problem in the so-called
\textit{reionization scenario} (e.g. Bullock et al. 2000; Benson
et al. 2002; Somerville 2002; Ricotti \& Gnedin 2005; Moore et al.
2006). The key assumption of this reionization scenario is that
only halos which acquire a significant amount of mass before the
redshift of reionization are able to form stars.
%{\it Please clarify.}
DSphs formed before the ionization era, are prevented from forming
stars by photoionization feedbacks (e.g. Babul \& Rees 1992; Quinn
et al. 1996; Weinberg et al. 1997; Navarro \& Steinmetz 1997).
%}

In Section \ref{sec:Seg1}, we summarize observational results and
inferences. In Section \ref{sec:surface}, the surface mass
densities of the stellar and gas discs are estimated according to
XLD model. Regarding Segue 1 as a dwarf disc galaxy, we construct
possible stellar equilibrium configurations in Section
\ref{sec:equilibrium,nogas}. In Section \ref{sec:equilibrium,gas},
we further assume that gas disc and magnetic field are also
present in Segue 1. For this configuration, we construct possible
equilibrium configurations and estimate magnetic field strength as
well as the resulting ratio $f$.

\section{Observations of dwarf galaxy Segue 1}\label{sec:Seg1}

Segue 1 does not appear spherical or bulge-like visually in figure
1 of Geha et al. (2009); it has a fairly long stretch with a
projected thickness. By this morphological appearance, Segue 1 is
most likely a dwarf disc galaxy almost edge-on to be consistent
with figure 1 of Geha et al. (2009).
%\ref{fig:GehaFig1}.
%\begin{figure}
%\begin{center}
%\includegraphics[width=14cm]{geha.png}
%\caption{}
%\label{fig:GehaFig1}
%\end{center}
%\end{figure}
In Table \ref{tab:resultsGeha}, observational results of Geha et
al. are summarized.
%{\bf Indicate their assumptions.}
%
%{\bf \newline
To determine the total dynamic mass $M^{(tot)}$ within $\sim$ 50
pc, they used two methods leading to two slightly different masses
(see Table \ref{tab:resultsGeha}). Instead of a disc galaxy, both
methods assume that Segue 1 is a relaxed, self-gravitating,
spherically symmetric system without rotation.

The first method assumes a
%isotropic
sphere where mass follows light. The density is described by
King's model (1966) in a virial equilibrium. The total mass is
determined according to Illingworth (1976) to be
$M^{(tot)}=167\beta^* r_c \big[\sigma^{(s)}\big]^2$, where
$\beta^*=8$ for typical dSphs (e.g. Mateo 1998),
$r_c=18.6^{+5}_{-3}\ \rmn{pc}$ is the core radius of King's
profile for Segue 1, and $\sigma^{(s)}$ is the mean stellar
velocity dispersion.

The second method is detailed in Strigari et al. (2008). The two
main assumptions used by Geha et al. (2009) are that the light
profile follows the observed Plummer profile with effective radius
$r_{\rm eff}=29\ \rmn{pc}$, and that the DM follows a
five-parameter density profile (Strigari et al. 2008). By
marginalizing over these parameters for the DM density profile,
the mass at any radius $r$ is determined.

Putman et al. (2008) noted that Segue 1 has little to almost no HI
gas, with an upper limit of $\sim 13\ \rmn{M_\odot}$ for the gas
mass. In Table \ref{tab:resultsGeha}, we include the corresponding
ratios $f$ between DM and baryon masses, or equivalently, the DM
potential to the baryon potential.

%{\bf Please be more specific and outline our model results clearly.
%\newline
%In Table \ref{tab:resultsGeha}, we still have no results of our
%calculations. The ratios $f=M^{(DM)}/M^{(d)}$ between the dark
%matter mass and disc mass are calculated by only using Geha's
%results of the total dynamical mass $M^{(tot)}$, the stellar mass
%$M^{(s)}$ and Putman's result of the upper limit of the gaseous
%mass $M^{(g)}$. Then, I used $M^{(d)}=M^{(s)}+M^{(g)}$ and dark
%matter mass $M^{(DM)}=M^{(tot)}-M^{(d)}$. }
%
%They suggest that Segue 1 (as the other gas-less dwarf galaxies in
%Figure 1 of Putman et al. 2008) once had a similar gas mass to the
%dwarf galaxies with gas, $10^{5-7}\ \rmn{M_\odot}$. The gas content
%is given to the Galactic disk.
\begin{table*}
%\centering
\caption{Key parameters of dwarf galaxy Segue 1 as inferred by
Geha et al. (2009).
%{\bf Please define notations clearly. \newline
The mean heliocentric radial velocity in the first row is the
radial velocity
%{\bf along the line of sight?}
from the Sun to Segue 1. Parameter $f$ in the 9th row is the ratio
between the DM mass $M^{(DM)}$ and the baryon mass
$M^{(d)}=M^{(s)}+M^{(g)}$ consisting of stellar mass $M^{(s)}$ and
gas mass $M^{(g)}$. The total dynamic mass $M^{(tot)}$ includes
$M^{(DM)}$ and $M^{(d)}$. The stellar mass $M^{(s)}$ is determined
by using a ratio of $M^{(s)}/L_V\sim 3$ (e.g. Maraston 2005)
between the mass and the luminosity in the absence of
%non-baryon
DM. In the 10th row, $f_{\rm min}$ and $f_{\rm max}$ are the
minimum and maximum values estimated for $f$ parameter. }
\begin{tabular}{|c|c|c|}
\hline
Variables & Values\\
\hline
\hline
Mean heliocentric radial velocity
%{\bf Meaning?}
& $206.4 \pm 1.3\ \rmn{km ~s^{-1}}$\\
\hline
Stellar velocity dispersion $\sigma^{(s)}$ & $4.3\pm 1.2\ \rmn{km~s^{-1}}$\\
\hline
Total stellar luminosity $L_V$ within 50 pc & $\sim 340\ \rmn {L_\odot}$ \\
\hline
Total stellar mass $M^{(s)}$ by using
$M^{(s)}/L_V \sim 3$ & $\sim 10^3 \ \rmn{M}_\odot$\\
\hline
& Mass-follow-light model results &
Two-component maximum-likelihood model results\\
\hline The calculated total dynamic mass $M^{(tot)}$ within 50 pc
& $4.5^{+4.7}_{-2.5}\times 10^5\ \rmn{M_\odot}$
& $8.7^{+13}_{-5.2} \times 10^5 \ \rmn{M_\odot}$ \\
\hline The resultant mass-to-light ratio $M^{(tot)}/L_V$ &
$1320^{+2680}_{-940}$ & $2440^{+1580}_{-1775}$ \\
\hline
\hline
$f=M^{(DM)}/M^{(d)}=\big[M^{(tot)}-M^{(d)}\big]/M^{(d)}$
& $443.2^{+464.0}_{-246.8}$ &  $ 857.8^{+1283.4}_{-513.3}$ \\
\hline
$f_{\rm min}$\qquad and\qquad $f_{\rm max}$ & 196.4 &  2142.2 \\
\hline
\hline
\end{tabular}
\label{tab:resultsGeha}
\end{table*}

\section{Estimates of surface mass densities}\label{sec:surface}

Our recent XLD model involves the rotational equilibrium of a
composite scale-free disc system embedded in an axisymmetric DM
halo; this composite disc system contains a thin stellar disc and
an isopedically magnetized thin gas disc.
Using the XLD model for a scale-free thin stellar disc in
cylindrical coordinates $(r,\theta,z)$
%Xiang-Gruess et al. (2009)
and a total stellar mass $M^{(s)}\sim 10^3\ \rmn{M_\odot}$ (Geha
et al. 2009) within $\sim 50\ \rmn{pc}$, the stellar surface mass
density profile in radius $r$ is
\begin{eqnarray}\label{stellarD}
\Sigma^{(s)}_0(r)=S^{(s)} r^{-2\beta-1}\ .
\end{eqnarray}
Here, the coefficient $S^{(s)}$ of eq. (12) in XLD
%Xiang-Gruess et al. (2009)
is determined by
\begin{eqnarray}
 M^{(s)}(50\ \rmn{pc})=\int_{0}^{50\ \rmn{pc}}
 \int_0^{2\pi }\Sigma_0^{(s)}(r) r dr d\theta\ ,\\
%10^3\ \rmn{M_\odot}=2 \pi S^{(s)} \frac{1}{(1-2 \beta)}
%\left[ r^{1-2\beta} \right]_0^ { 50\ \rmn{pc}}\ ,\\
S^{(s)}=\frac{(1-2\beta)10^3\ \rmn{M_\odot} }{2 \pi \left[
r^{1-2\beta} \right]_0 ^ { 50\ \rmn{pc}}}\ .
\end{eqnarray}
For the valid range of scaling index $\beta=(0,\ 1/2)$, we
estimate $S^{(s)}\in (2.06 \times 10^{15},\ 2.5\times 10^{33})\
\rmn{g\ cm^{2\beta-1}}$ in expression (\ref{stellarD}).
%The stellar surface mass density is
%given by
%\begin{eqnarray}
%\Sigma^{(s)}_0(r)=S^{(s)} r^{-2\beta-1}\ .
%\end{eqnarray}

For a thin gas disc with a gas surface mass density of
\begin{eqnarray}
\Sigma^{(g)}_0(r)=S^{(g)} r^{-2\beta-1}\
\end{eqnarray}
%and  the same procedure can be done by
%assuming a scale-free thin gas disc
and a total gas mass of $M^{(g)}\lsim 13\ \rmn{M_\odot}$ within
$\sim 50\ \rmn{pc}$, the constant coefficient $S^{(g)}$ is
similarly estimated by
\begin{eqnarray}\label{gasD}
 M^{(g)}(50\ \rmn{pc})=\int_{0}^{50\ \rmn{pc}}\int_0^{2\pi }
 \Sigma_0^{(g)}(r) r dr d\theta\ ,\\
%13\ \rmn{M_\odot}=2 \pi S^{(g)} \frac{1}{(1-2 \beta)}
%\left[ r^{1-2\beta}\right]_0^ { 50\ \rmn{pc}}\ ,\\
S^{(g)}=\frac{(1-2\beta) 13\ \rmn{M_\odot} } {2 \pi \left[
r^{1-2\beta} \right]_0^{ 50\ \rmn{pc}}}\ .\label{eq:Sg}
\end{eqnarray}
For the same range $\beta=(0,\ 1/2)$, eq (\ref{eq:Sg}) then gives
the corresponding $S^{(g)}\in (2.68\times 10^{13},\ 3.26\times
10^{31})\ \rmn{g\ cm^{2\beta-1}}$.
%The gas surface mass density is given by
%\begin{eqnarray}
%\Sigma^{(g)}_0(r)=S^{(g)} r^{-2\beta-1}\ .
%\end{eqnarray}

The ratio of the two disc surface mass densities is then
\begin{eqnarray}
\delta_0=\frac{\Sigma^{(g)}_0(r)}{\Sigma^{(s)}_0(r)}
=\frac{S^{(g)} }{ S^{(s)}}\sim\frac{13\ \rmn{ M_\odot}}{\
10^3\rmn{ M_\odot}} =0.013\ .
\end{eqnarray}
This ratio $\delta_0$ characterizes the global evolution of a
dwarf disc galaxy as stars form out of the gaseous interstellar
medium (ISM).

\section{Equilibrium states without gas disc}\label{sec:equilibrium,nogas}

For a single thin stellar disc without gas disc, the equilibrium
is sustained by the radial momentum balance (see XLD), viz.
\begin{eqnarray} \label{eq:nogas}
 \big[v^{(s)}_{\theta 0}\big]^2+\big[a^{(s)}\big]^2
 (2\beta+1)=2\beta r \rmn{G} Y_0(\beta) \Sigma^{(s)}_0(1+f)\ ,
\end{eqnarray}
where $v^{(s)}_{\theta 0}$ is the stellar disc rotation speed,
$a^{(s)}$ is the stellar velocity dispersion, $\rmn{G}=6.67\times
10^{-8}$ cm$^3$g$^{-1}$s$^{-2}$ is the gravitational constant, and
$Y_0(\beta)$ is related to the Gamma functions $\Gamma (z)$ by
\begin{equation}\label{Y0def}
Y_0(\beta)\equiv\frac{\pi\Gamma(1/2-\beta)
\Gamma(\beta)}{\Gamma(1-\beta)\Gamma(1/2+\beta)}\
\end{equation}
%Equation (\ref{eq:nogas}) is derived from a vanishing stellar
%radial velocity $v_\rmn{r0}^{(s)}=0$ and the radial momentum
%equation
(see eq (26) of XLD).
%Xiang-Gruess et al. 2009).
Without $v^{(s)}_{\theta 0}$, the corresponding ratio $f_{\rm
min}^{(s)}$ between the DM potential $\overline{\Phi}_0$ and
baryon mass potential $\Phi^{(s)}_0$ can be determined from eq
(\ref{eq:nogas}) within the following range of
\begin{eqnarray}
%\!\!\!\!\!\!\!\!\!
f_{\rm min}^{(s)} = \frac{\big[a^{(s)}\big]^2
 (2\beta+1)}{2\beta r GY_0(\beta) \Sigma^{(s)}_0}-1 =
 \left \{ \begin{array}{c}
                         85.7 \ \rmn{for} \ \beta=0.49 \ ,  \\
             208.5 \ \rmn{for} \ \beta=0.01 \ ,
                          \end{array} \right. \label{eq:f_min^s}
\end{eqnarray}
where we adopt $a^{(s)}\sim 4.3\ \rmn{km\ s}^{-1}$ at $r=10\
\rmn{pc}$. This range of $f_{\rm min}^{(s)}$ shifts upwards for a
larger $a^{(s)}$.
%{\bf You mean $r=10\ \rmn{pc}$? \newline
%Yes, I meant pc, I have corrected it.}
Scaling index $\beta$ has a theoretically allowed range of $\beta
\in (0,\ 0.5)$. This range of lower limits for $f_{\rm min}^{(s)}$
already indicates that $f$ must be unusually large for
%the dwarf galaxy
Segue 1, as the stellar disc itself is not sufficiently massive to
counteract the `stellar pressure' mimicked by the stellar velocity
dispersion $a^{(s)}$. For $v^{(s)}_{\theta 0}\neq 0$, the ratio
$f$ should be even higher by eqn (\ref{eq:nogas}).

\begin{figure}
\begin{center}
%\begin{tabular}{c}
\resizebox{70mm}{!}{\includegraphics{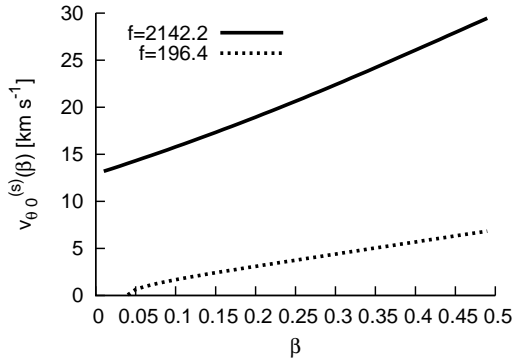}}
%\resizebox{70mm}{!}{\includegraphics{v_beta_f_max.eps}} \\
%\resizebox{70mm}{!}{\includegraphics{v_beta_f_min.eps}}
%\end{tabular}
\caption{Displayed here are the stellar disc rotation velocities
$v_{\theta 0}^{s}$ in unit of $\rmn{km\ s}^{-1}$ for two given
potential ratio limits $f_{\rm max,obs}=2142.2$ (upper solid
curve) and $f_{\rm min,obs}=196.4$ (lower dotted curve) with a
stellar velocity dispersion $a^{(s)}(10\ \rmn{pc})=4.3\ \rmn{km\
s}^{-1}$ as functions of the disc scaling index $\beta$ parameter.
%{\bf Adjust subscripts?}
} \label{fig:v_beta}
\end{center}
\end{figure}
For the observed stellar velocity dispersion $a^{(s)}$ and
$f={\overline{\Phi}_0}/{\Phi^{(s)}_0}$ between the DM potential
$\overline{\Phi}_0$ and the baryon potential $\Phi^{(s)}_0$, we
determine the necessary disc rotation speed $v^{(s)}_{\theta 0}$
using eq (\ref{eq:nogas}).
% (\ref{eq:v}).
In Fig. \ref{fig:v_beta}, we show the disc rotation speeds for the
upper limit $f_{\rm max,obs}=2142.2$ and the lower limit $f_{\rm
min,obs}=196.4$ inferred by Geha et al. (2009) as a function of
our disc scaling index $\beta$.
%\begin{eqnarray}
%v^{(s)}_{\theta 0}=\Big\{2\beta r GY_0(\beta)\Sigma^{(s)}_0
%(1+f)-\big[a^{(s)}\big]^2(2\beta+1)\Big\}^{1/2}\ . \label{eq:v}
%\end{eqnarray}
For $f_{\rm max,obs}=2142.2$, the equilibria for all allowed
$\beta$ can persist for $v_{\theta 0}^{(s)}\neq 0$. For $f_{\rm
min,obs}=196.4$ and $\beta\cong 0.04$, $v_{\theta 0}^{(s)}=0$
allows an equilibrium, whereas for all $\beta >0.04$, one infers
$v_{\theta 0}^{(s)}\neq 0$ for the stellar disc.

For a single stellar disc without rotation embedded in an
axisymmetric DM halo, the minimum $f$ falls in the range $\sim
86-209$ depending on $\beta$ value. This trend grossly agrees with
recent observations that Segue 1 is a DM dominated dwarf galaxy.
Nevertheless, for $f$ approaches several hundreds or even $\sim
2000$ as shown in Geha et al. (2009), a stellar disc rotational
speed is necessarily required as XLD model cannot accommodate such
large amounts of DM by $a^{(s)}\sim 4.3\ \rmn{km\ s}^{-1}$ alone.
%only using the stellar velocity dispersion.
The corresponding stellar disc rotation speeds estimated from
$f_{\rm min,obs}$ and $f_{\rm max,obs}$ (Fig. \ref{fig:v_beta})
are much slower than those typical of disc galaxies, which are
$\gsim 150 \ \rmn{km\ s}^{-1}$. While $v^{(s)}_{\theta 0}$
inferred from $f_{\rm min,obs}$ appears fairly small for a disc
galaxy with a stellar velocity dispersion of $a^{(s)}(10\
\rmn{pc})\sim 4.3\ \rmn{km\ s}^{-1}$, a speed $v^{(s)}_{\theta 0}$
inferred from $f_{\rm max,obs}$ of $\sim 30\ \rmn{km\ s}^{-1}$ may
be plausible for dwarf disc galaxy Segue 1.

In conclusion, by grossly fitting Segue 1 with XLD dynamic model,
%(Xiang-Gruess et al. 2009),
we estimate a range of $f\sim 86-209$ for a composite disc system
without disc rotation. For values of $f$ approaching $\sim 2000$,
a disc rotation speed up to $v_{\theta 0}^{(s)}\sim 30\ \rmn{km\
s}^{-1}$ is necessary. A more precise determination of $f$ is thus
highly desirable in order to estimate the disc rotation speed and
the applicability of XLD model. Thereby, the method of only using
the stellar velocity dispersion $a^{(s)}$ is not adequate due to
the contribution from $v_{\theta 0}^{(s)}\neq 0$ to $f$ (see eq
\ref{eq:nogas}). At this stage, our model is able to produce large
values up to $\sim 200$ for $f$ even without stellar disc
rotation. This generally agrees with the lower results of Geha et
al. (2009).

\section{Equilibrium disc system in the presence \quad\ \
of an isopedically magnetized gas disc}\label{sec:equilibrium,gas}

By including a thin scale-free isopedically magnetized gas disc
according to XLD,
%Xiang-Gruess et al. (2009),
the equilibrium state involves two coupled radial momentum
balances for two discs embedded in a DM halo, viz.
\begin{eqnarray}\label{eq:1.a}
\big[v^{(s)}_{\theta 0}\big]^2+\big[a^{(s)}\big]^2(2\beta+1)
\qquad\qquad\qquad\qquad\qquad \\ \nonumber =2\beta r GY_0(\beta)
\big[\Sigma^{(s)}_0+\Sigma^{(g)}_0\big](1+f)\ ,\\
\label{eq:2.a}
\big[v^{(g)}_{\theta 0}\big]^2
+\Theta\big[a^{(g)}\big]^2(2\beta+1) \qquad\qquad\qquad\qquad
\\ \nonumber =2\beta rGY_0(\beta)
\left\lbrace\big[\Sigma^{(s)}_0+\Sigma^{(g)}_0\big]
(1+f)-(1-\epsilon)\Sigma^{(g)}_0\right\rbrace\ ,
\end{eqnarray}
where $v^{(g)}_{\theta 0}$ is the gas disc rotation speed,
$a^{(g)}$ is the gas sound speed, $\Theta$ and $\epsilon$ are
functions of the isopedic magnetic field strength $B_z$ with
$\lambda=2\pi \rmn{G}^{1/2}\Sigma^{(g)}_0/B_z$, $\ \eta= {\beta
Y_0(\beta)}/{\pi}$, $\ \epsilon = 1-\lambda^{-2}$, $\
\Theta=1+{(1+\eta^2)}/{(\hat{\lambda}^2+\eta^2)}$ and $\
\hat{\lambda}=\big(1+\delta_0^{-1}\big)\lambda$ (see Lou \& Wu
2005 and XLD for more details).

For this composite disc system embedded in a massive DM halo, the
ratio $f$ is defined as
$f={\overline{\Phi}_0}/{\big[\Phi^{(s)}_0+\Phi^{(g)}_0 \big]}$.
Taking the total gas mass inside $\sim 50$ pc as $M^{(g)}(50\
\rmn{pc})\sim 13\ \rmn{M_\odot}$, we now explore consequences of
radial momentum balances (\ref{eq:1.a}) and (\ref{eq:2.a}).
%By using the same rotational velocity for the gaseous disc
%$v^{(g)}_{\theta 0}=v^{(s)}_{\theta 0}$ and the same
%two limits for $f$, only the gaseous velocity dispersion
%$a^{(g)}$ and the two constants $\Theta$ and $\epsilon$
%characterizing the magnetic field strength are unknown.

\subsection{Composite equilibria without disc rotations}

We first assume that both disc rotation speeds $v_{\theta
0}^{(s)}$ and $v_{\theta 0}^{(g)}$ vanish as in Geha et al.
(2009).
%In this way, we can derive the required magnetic field
%strength for a given gaseous sound speed $a^{(g)}$.
By eqn (\ref{eq:1.a}) with $v^{(s)}_{\theta 0}=0$, the ratio
$f^{(s)}_{\rm min}$ for the stellar disc is allowed in an
equilibrium. This $f^{(s)}_{\rm min}=f^{(g)}_{\rm min}$ must also
be set in eq (\ref{eq:2.a}) as for a composite system $f$ is the
same for stellar and gas discs. By also using $v^{(g)}_{\theta
0}=0$ and a certain gas sound speed $a^{(g)}$, the corresponding
magnetic field strength characterized by $\Theta$ and $\epsilon$
parameters is found. In this way, we can construct an equilibrium
for the composite system of a stellar and a magnetized gas disc
without rotation and infer a magnetic field strength. We calculate
below the corresponding isopedic magnetic field strength $B_z$ in
order to discuss its observational diagnostics (Lou \& Fan 2003;
Lou \& Wu 2005; Wu \& Lou 2006).

For the stellar disc, $f_{\rm min}^{(s)}$ for $v_{\theta
0}^{(s)}=0\ \rmn{km\ s}^{-1}$ is given by
\begin{eqnarray}
f^{(s)}_{\rm min} = \frac{\big[a^{(s)}\big]^2 (2\beta+1) }{2\beta
r GY_0(\beta) \big[\Sigma^{(s)}_0+\Sigma^{(g)}_0\big]}-1
\qquad\qquad\qquad\qquad \\ \nonumber = \left \{ \begin{array}{c}
                         84.6 \ \rmn{for} \ \beta=0.49 \ ,  \\
             205.8 \ \rmn{for} \ \beta=0.01\ .
                          \end{array} \right. \label{eq:f^s}
\end{eqnarray}
For the magnetized gas disc, $f_{\rm min}^{(g)}$ for $v_{\theta
0}^{(g)}=0\ \rmn{km\ s}^{-1}$ is given by
\begin{eqnarray}
f^{(g)}_{\rm min}=\frac{\Theta \big[a^{(g)}\big]^2 (2\beta+1)
}{2\beta r GY_0(\beta)\big[\Sigma^{(s)}_0 +\Sigma^{(g)}_0\big]} +
\frac{(1-\epsilon)\Sigma^{(g)}_0}
{\big[\Sigma^{(s)}_0+\Sigma^{(g)}_0\big]}-1\ .\label{eq:f^g}
\end{eqnarray}
For a gas disc sound speed $a^{(g)}(10\rmn{pc})=0.5\ \rmn{km\
s}^{-1}$, we find $f^{(g)}_{\rm min}=84.6$ for
%$\lambda=0.0124126 $ or
$B_z(1\rmn{pc})=1.05\rmn{\mu G}$ with $\beta=0.49$ or
$f^{(g)}_{\rm min}=205.8$ for
%$\lambda= $ or
$B_z(1\rmn{pc})=1.87\rmn{\mu G}$ with $\beta=0.01$.
%{\bf 1 pc or 10 pc?}
These magnetic field strengths are commonly inferred in many disc
spiral galaxies (e.g. Fan \& Lou 1996; Lou \& Fan 1998) and should
be observationally searched for Segue 1 as a test or a constraint
of XLD model. We thus advance a model configuration for Segue 1 by
using a composite system containing stellar and magnetized gas
disc components embedded in a massive axisymmetric DM halo (XLD).

For zero magnetic field with $\epsilon\to 1$ and $\Theta\to 1$,
the gas disc must rotate in order to have the same $f$ as the
non-rotating stellar disc. For $\beta=0.49$, the ratio $f_{\rm
min}^{(s)}$ of a non-rotating stellar disc is 84.6; for
$a^{(g)}=0.5\ \rmn{km\ s}^{-1}$, the corresponding $v_{\theta
0}^{(g)}$ is 6 km s$^{-1}$ by eq (\ref{eq:2.a}). For $\beta=0.01$,
the ratio $f_{\rm min}^{(s)}$ of a non-rotating stellar disc is
205.8 and the $v_{\theta 0}^{(g)}$ is 4.3 km s$^{-1}$ for the same
$a^{(g)}$.

\subsection{Influence of an isopedic magnetic field in the ISM disc}

\begin{figure}
\begin{center}
\begin{tabular}{cc}
\resizebox{70mm}{!}{\includegraphics{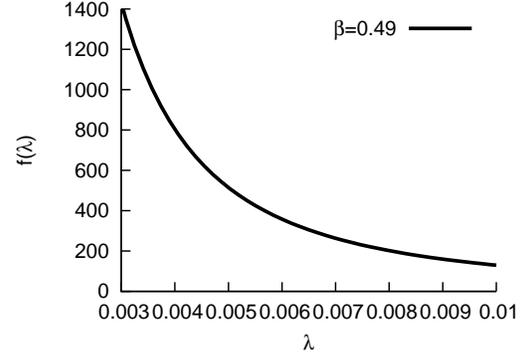}}
\end{tabular}
\caption{Shown here is the dependence of the gravitational
potential ratio $f$ (i.e. the potential of DM over that of baryon
disc mass) on the dimensionless magnetic parameter $\lambda=2\pi
G^{1/2}\Sigma_0^{(g)}/B_z$ for the gaseous ISM disc. The magnetic
field strength range of $B_z\in (1,\ 5)\ \rmn{\mu G}$ is adopted
given other inferred parameters of dwarf disc galaxy Segue 1.
%{\bf The red colour is not needed.\newline
%I made a new plot with black curve}
}\label{fig:f_l}
\end{center}
\end{figure}

%The equilibrium state is characterized by a certain $f$ and certain (different) rotational velocities.
%For the calculation of lower limits of $f$, we set $\big[v^{(s)}_{\theta 0}\big]^2=\big[v^{(g)}_{\theta 0}\big]^2=0$ in order to get the total lower limit
%$f_{min}^{(tot)}$ of the whole composite system which is
%\begin{eqnarray}
% f_{min}^{(tot)}= \left \{ \begin{array}{c}
%                         f_{min}^{(s)} \ \rmn{for} \ f_{min}^{(s)} \geq f_{min}^{(g)} \ ,  \\
%            f_{min}^{(g)} \ \rmn{for} \ f_{min}^{(g)}>f_{min}^{(s)}  \ .
%                          \end{array} \right. \label{eq:f_min,tot}
%\end{eqnarray}
We now examine eqn (\ref{eq:f^g}) with the emphasis on the
influence of magnetic field on $f$ for Segue 1. By assuming
$v_{\theta 0}^{(g)}=0$, we plot the corresponding $f$ ratio for
$\beta=0.49$, $a^{(g)}=(10\ \rmn{pc})=0.5\ \rmn{km\ s}^{-1}$ and
for $B_z \in (1,\ 5)\ \rmn{\mu G}$ at $r=1$ pc or $\lambda \in
(0.01,\ 0.003)$ in Fig. \ref{fig:f_l}. One should note the inverse
proportionality of $\lambda$ to $B_z$, i.e. a stronger magnetic
field corresponds to a smaller $\lambda$ and vice versa.
For magnetic field strengths up to $10\ \rmn{\mu G}$, %at $r=1\ \rmn{pc}$
the corresponding $f_{\rm min}^{(g)}$ grows to very large values.
It is remarkable that relatively weak magnetic fields of a few
$\rmn{\mu G}$ can cause a rapid increase of $f_{\rm min}^{(g)}$
(XLD). Thereby, $f$ may grow into a range for which the stellar
disc cannot maintain an equilibrium without a rotation speed as
discussed above. For Segue 1, if it is possible in the near future
to estimate the magnetic field strength via synchrotron radio
emissions, then we could also infer the stellar disc rotation
speed.

Physically, the composite model of a magnetized gas disc may thus
correspond to a large $f$ ratio without the requirement of gas
disc rotation, in contrast to the stellar disc where we can only
reach $f\sim 200$ without disc rotation given the currently
estimated stellar velocity dispersion.

For zero magnetic field on the other hand, we find a lower limit
of $f_{\rm min}^{(g)}\sim 14$ which lies well beneath the lower
limit $f_{\rm min}^{(s)}$ found for the stellar disc. In this
case, the gas disc must rotate in order to reach the high $f$
determined by the stellar disc. The gas disc rotation speeds
$v_{\theta 0}^{(g)}$ for the two ratios $f_{\rm min}^{(s)}$ of
$\beta=0.01$ and $\beta=0.49$ are calculated for Segue 1 at the
end of the last subsection.

Therefore for weak or no magnetic fields, the gas disc, if there
is indeed one in Segue 1, must then rotate. For magnetic field
strengths of at least $\sim 2\ \rmn{\mu G}$, the ratio $f_{\rm
min}^{(g)}$ given by the gas disc is larger than the ratio $f_{\rm
min}^{(s)}$ of the stellar disc without rotation. In this case,
the stellar disc must be in rotation. For magnetic field strengths
in the range of $[2,\ 10]\ \rmn{\mu G}$ at radius $1\ \rmn{pc}$,
the corresponding stellar disc rotation speeds $v_{\theta
0}^{(s)}$ for $\beta=0.49$ are $[9.7,\ 45.5]\ \rmn{ km\ s}^{-1}$,
while for $\beta=0.01$, we have $v_{\theta 0}^{(s)}=[1.6,\ 22.5]\
\rmn{km\ s}^{-1}$.
By estimating the magnetic field strength and/or the $f$ ratio for
Segue 1, we can infer which disc must or can be in rotation.

\section{Discussion and speculations}

At this stage of investigation, two aspects of Segue 1 still
remain uncertain in terms of observations. (i) It is not sure
whether Segue 1 has a gas disc or not; only an upper mass limit
can be estimated at present. (ii) While no rotation is inferred so
far for Segue 1 (Geha et al. 2009), a disc rotation still cannot
be excluded definitely due to the small number of stars sampled so
far.

Because of these uncertainties, we performed several calculations
for two different cases, i.e. without and with gas disc. By
assuming Segue 1 in an equilibrium, we test the hypothesis that
the baryon component is distributed in a thin disc whereas the DM
is in an axisymmetric halo surrounding Segue 1. We mainly use the
results for the stellar disc velocity dispersion $a^{(s)}$ and the
upper and lower limits $f_{\rm max,obs}$ and $f_{\rm min,obs}$
inferred by Geha et al. (2009).

In the first study, we assume no gas disc in the dwarf disc galaxy
Segue 1. The stellar disc mass is set to be $\sim 10^3\
\rmn{M_\odot}$ inside $r\sim 50\ \rmn{pc}$ and the stellar
velocity dispersion is mimicked as the sound speed of the stellar
disc. Without rotation, we are able to derive the corresponding
ratio $f^{(s)}_{\rm min}$ as a function of scaling index $\beta$.
By comparing our $f^{(s)}$ with the lower and upper limits $f_{\rm
min,obs}$ and $f_{\rm max,obs}$ found by Geha et al. (2009), we
find that for $\beta\geq 0.04$, a disc rotation is necessary in
order to reach the lower limit $f_{\rm min,obs}$. For the upper
limit $f_{\rm max,obs}$, for all theoretical allowed $\beta$, a
disc rotation speed is needed. Our solutions for $f$ ratio hence
agree with $f_{\rm min,obs}$ and $f_{\rm max,obs}$, if disc
rotations are allowed. If it is found observationally that no disc
rotation is present, then our composite model provides a range of
$(86,\ 209)$ for $f$. The large amount of DM is required due to
the large stellar velocity dispersion representing an `effective
pressure' in the stellar disc compared to the low stellar surface
mass density. Since the stellar disc is not capable to counteract
the strong disc `pressure', the massive DM halo is required. Our
upper limit $f_{\rm max}^{(s)}=209$ is much smaller than $f_{\rm
max,obs} \cong 2142$. If a $f_{\rm max}^{(s)}=2142$ is indeed
necessary, then a disc rotation speed of $v_{\theta 0}^{(s)}\cong
30\ \rmn{km\ s}^{-1}$ (for $\beta=0.49$) and $v_{\theta
0}^{(s)}\cong 13\ \rmn{km\ s}^{-1}$ (for $\beta=0.01$) is needed.
Our solutions for $f$ hence agree with inferred $f_{\rm min,obs}$
and $f_{\rm max,obs}$ of Geha et al. (2009), only if disc
rotations are allowed. In short, our disc model analysis suggests
that a determination of $f$ with only the estimated velocity
dispersion is insufficient.

In the second study, we assumed a much less massive gas disc
component in Segue 1. Without magnetic fields, the DM amount is
mainly associated with the stellar disc and the gas disc is
expected to be in rotation. For an isopedic magnetic field of at
least $\sim 2\ \rmn{\mu G}$, the influence of the magnetic field
is so strong that the value of $f_{\rm min}^{(g)}$ exceeds the
$f_{\rm min}^{(s)}$, i.e. the stellar disc has to rotate in order
to reach $f_{\rm min}^{(g)}$. With increasing magnetic field
strengths, $f_{\rm min}^{(g)}$ increases rapidly in order to
counteract the magnetic Lorentz force in the gaseous ISM disc.

In terms of dynamics, we offer a plausible theoretical explanation
for the large amount of DM in the dwarf disc galaxy Segue 1. Our
main assumption is a baryon disc embedded in an axisymmetric DM
halo to sustain a global equilibrium.

The two major forces which the DM halo has to confine are the
strong effective pressure in the stellar disc and in the presence
of gas disc and isopedic magnetic field therein, the magnetic
Lorentz force in the gaseous ISM disc. Without rotation, these
forces are essential to counterbalance the self-gravity of a
massive DM halo.

An important conclusion of this study is, that a large amount of
DM may not completely correspond to the stellar dynamics in Segue
1 type dwarf disc galaxies, but might also correspond to magnetic
fields embedded in gas discs of such dwarf galaxies.

As stellar velocity dispersions alone cannot determine $f$
uniquely,
%{\bf
we discuss potentially possible independent means of
estimating DM halo mass of Segue 1, e.g. gravitational lensing.
%and dynamic inferences.
%1. Gravitational lensing:
First mentioned by Einstein (1936) and Zwicky (1937a,b), the
gravitational lensing effect was detected by Walsh et al. (1979).
Recently, the possibility of detecting DM halos of dwarf galaxies
was studied by Zackrisson et al. (2008), Riehm et al. (2009) and
Zackrisson \& Riehm (2009). Theoretically, gravitational lensing
effects can be utilized to determine the DM halos of dwarf
galaxies. Practically, the expected effects of DM halos are still
too weak to be observed. But this method is a possible method for
the future to estimate masses of dwarf galaxies.
%}

%2. Dynamical inferences: According to Piatek et al. (2002), the
%proper motion of a dwarf galaxy can be determined with the help of
%images taken at different times. Piatek et al. (2002) showed, that
%with the galactocentric velocity given, a lower limit for the mass
%of the MW can be determined. But, I actually do not know how one
%can determine the mass of the satellite dwarf galaxy by using its
%rotation around the MW. The only relation we can use is
%centrifugal force balances gravitational force and this relation
%only gives the mass of the MW. }

Another point
%that we have discussed
is that in the gas disc, a magnetic field could be the counterpart
of the massive DM halo preventing the necessity of gas disc
rotation. For this study, a determination of magnetic field is
crucial for Segue 1.
%
%{\bf
Blasi et al. (2003) discussed synchrotron radio emissions from
galactic satellites. In DM dominated dwarf galaxies, high-energy
electrons and positrons may be expected as by-products of
high-energy photons (from the $\pi^{\pm}$ decay chains) which
might be on their part products of DM annihilations. In the
presence of magnetic fields threading through a gas disc, these
high-energy electrons and positrons may be revealed via
synchrotron radio emissions. Due to its massive DM halo, Segue 1
could be a promising candidate for high-energy electrons and
positrons. Along this line, the question whether there is a
magnetic field or not can be tested observationally for Segue 1 by
searching for extended synchrotron radio emissions (Lou \& Fan
2003). In reality, other magnetic field configurations are also
possible (e.g. Lou \& Zou 2004, 2006).

Finally, the issue, whether Segue 1 is a dwarf galaxy or a stellar
cluster, is still not answered definitively. The XLD model adapted
here for Segue 1 can be applied to other dwarf disc galaxies or
flat stellar clusters. Dwarf galaxies of similar stellar and/or
gas masses and velocity dispersions should lead to similar
conclusions according to our hydrodynamic model considerations.
%}

\section*{Acknowledgments}

This research has been supported in part by Deutscher Akademischer
Austauschdienst (DAAD; German Academic Exchange Service). This
research was supported in part
%by the ASCI Center for Astrophysical Thermonuclear Flashes at the
%University of Chicago, by the Special Funds for Major State Basic
%Science Research Projects of China,
by the Tsinghua Centre for Astrophysics,
%by the Collaborative Research Fund from the National Science
%Foundation of China (NSFC) for Young Outstanding Overseas
%Chinese Scholars (NSFC 10028306) at the National Astronomical
%Observatories, Chinese Academy of Sciences,
by NSFC grants 10373009 and 10533020 at the Tsinghua University,
and by the SRFDP 20050003088 and 200800030071, and the Yangtze
Endowment from the Ministry of Education at the Tsinghua
University.
%The hospitality
%of Institut f\"ur Theoretische Physik und Astrophysik
%der Christian-Albrechts-Universit\"at Kiel Germany and
%of International Center for Relativistic Astrophysics Network
%(ICRANet) Pescara, Italy is gratefully acknowledged.

%\bsp ``This paper has been produced using the ...''
\label{lastpage}

\begin{thebibliography}{07}
\bibitem{Ba} Babul A., Rees M. J., 1992, MNRAS, 255, 346
%ionization feedback

%\bibitem{Bal} Baltz E. A., Wai L., 2004, Phys. Rev.,
%D70, 23512 %%synchr. emission in dwarf galaxies

\bibitem{Bel} Belokurov V., et al. 2007, ApJ, 654, 897

\bibitem{Be} Benson A. J., et al.
%Frenk C. S., Lacey C. G., Baugh C. M., Cole S.,
2002, MNRAS, 333, 177
%reionization scenario

\bibitem{Bla} Blasi P., Olinto A. V., Tyler C., 2003,
Astropart. Phys., 18, 649
%synchr. emission in dwarf galaxies

\bibitem{Bu} Bullock J. S., et al.
%Kravtsov A. V., Weinberg D. H.,
2000, ApJ, 539, 517
%reionization scenario

\bibitem{Ei} Einstein A., 1936, Science, 84, 506
%gravitational lensing

%\bibitem{Ess} Essig R., Seghal N., Strigari L. E.,
%2009, astro-ph/09024750 %gamma-ray of Segue 1

\bibitem{FanLou1996} Fan Z. H., Lou Y.-Q., 1996, Nat,
383, 800
%-802; ``Origin of the Magnetic Spiral
%Arms in the Galaxy NGC6946"

\bibitem{Ge} Geha M., et al.
%Willman B., Simon J. D., Strigari L. E.,
%Kirby E. N., Law D. R., Strader J.,
2009, ApJ, 692, 1464
% Segue 1 result

%\bibitem{Go} Gould A., Bennett D. P., Alves D. R., 2004,
%ApJ, 614, 404 % microlensing, mass determination of lens

\bibitem{Il} Illingworth G., 1976, ApJ, 204, 73
%1. method of mass determination

\bibitem{Kau} Kauffmann G., et al.
%White S. D. M., Guiderdoni B.,
1993, MNRAS, 264, 201
%missing satellite problem

\bibitem{Kin} King I. R., 1966, AJ, 71, 64
% King model, 1. method of mass determination

\bibitem{Kly} Klypin A., et al.
%Kravtsov A. V., Valenzuela O., Prada F.,
1999, ApJ, 522, 82
%Where Are the Missing Galactic Satellites?
%related to the missing satellite problem
%
%According to the hierarchical clustering scenario, galaxies are
%assembled by merging and accretion of numerous satellites of
%different sizes and masses. This ongoing process is not 100%
%efficient in destroying all of the accreted satellites, as
%evidenced by the satellites of our Galaxy and of M31. Using
%published data, we have compiled the circular velocity (V_circ)
%distribution function (VDF) of galaxy satellites in the Local
%Group. We find that within the volumes of radius of 570 kpc
%(400 h^-1 kpc assuming the Hubble constant h=0.7) centered
%on the Milky Way and Andromeda, the average VDF is roughly
%approximated as n(>V_circ)~55+/-11(V_circ/10 km s^-1)^-1.4+/-0.4
%h^3 Mpc^-3 for V_circ in the range ~10-70 km s^-1. The observed
%VDF is compared with results of high-resolution cosmological
%simulations. We find that the VDF in models is very different from
%the observed one: n(>V_circ)~1200(V_circ/10 km s^-1)^-2.75 h^3
%Mpc^-3. Cosmological models thus predict that a halo the size
%of our Galaxy should have about 50 dark matter satellites with
%circular velocity greater than 20 km s^-1 and mass greater than
%3x10^8 M_solar within a 570 kpc radius. This number is
%significantly higher than the approximately dozen satellites
%actually observed around our Galaxy. The difference is even larger
%if we consider the abundance of satellites in simulated galaxy
%groups similar to the Local Group. The models predict ~300
%satellites inside a 1.5 Mpc radius, while only ~40 satellites are
%observed in the Local Group. The observed and predicted VDFs cross
%at ~50 km s^-1, indicating that the predicted abundance of
%satellites with V_circ>~50 km s^-1 is in reasonably good agreement
%with observations. We conclude, therefore, that unless a large
%fraction of the Local Group satellites has been missed in
%observations, there is a dramatic discrepancy between observations
%and hierarchical models, regardless of the model parameters. We
%discuss several possible explanations for this discrepancy
%including identification of some satellites with the high-velocity
%clouds observed in the Local Group and the existence of dark
%satellites that failed to accrete gas and form stars either
%because of the expulsion of gas in the supernovae-driven winds or
%because of gas heating by the intergalactic ionizing background.

\bibitem{LouFan1998}Lou Y.-Q., Fan Z. H., 1998, ApJ, 493, 102
%--120; ``Fast and Slow Density Waves
%in Magnetized Spiral Galaxies"

\bibitem{LouFan2003}Lou Y.-Q., Fan Z. H., 2003, MNRAS, 341, 909
%--926; ``Magnetohydrodynamic Density Waves in a Composite
%Disk System of Interstellar Medium and Cosmic-Ray Gas"

\bibitem{LouWu2005}Lou Y.-Q., Wu Y., 2005, MNRAS, 364, 475
%``Global Perturbation Configurations in a Composite
%Disc System with an Isopedic Magnetic Field",
%{\it Monthly Notices of the Royal Astronomical
%Society}, {\bf 364}, 475-502, 2005. astro-ph/0508601

\bibitem{LouZou2004}Lou Y.-Q., Zou Y., 2004, MNRAS, 350, 1220
%-1252,
%LouZou04.pdf received on May 21, 2004 (Friday)
%during a trip to Shanghai Observatory for Liao Xin-Hao group review.
%in directory /root/ZouYue/Galley01
%/Windows/diskd/ZouYue/Galley01/LouZou04.pdf
%directory ZouYue/Draft1/FigureR
%LaTex file Composite_MSIDr3.tex;
%on oddjob, thca(166.111.26.19), aire(166.111.26.56),
%louyq@166.111.26.61,
%rm3211(166.111.26.63).
%Email acceptance: December 5, 2003
%MNRAS manuscript reference number: MD1154
%Reference number: astro-ph/0312082

\bibitem{LouZou2006}Lou Y.-Q., Zou Y., 2006, MNRAS, 366, 1037
%``Axisymmetric Stability Criteria for a Composite System of
%Stellar \break and Magnetized Gaseous Singular Isothermal Discs",
%Lou Y.-Q., Zou Y., {\it Monthly Notices of the Royal Astronomical
%Society}, {\bf 366}, 1037-1049, 2006; astro-ph/0511348

\bibitem{Mar} Maraston C., 2005, MNRAS, 362, 799
%M/L_V=3 in absence of DM

\bibitem{Ma} Martin N. F., de Jong J. T. A., Rix H.-W., 2008, ApJ, 684, 1075

\bibitem{Mateo93} Mateo M., et al.
%Olszewski E. W., Pryor C., Welch D. L., Fischer P.,
1993, AJ, 105, 510

\bibitem{Mt} Mateo M. L., 1998, ARA \& A, 36, 435
    %beta of 1. method of mass determination

\bibitem{Mo} Moore B., et al.
  %Diemand J., Madau P., Zemp M., Stadel J.,
  2006, MNRAS, 368, 563
  %reionization scenario

\bibitem{Moo} Moore B., et al.
   %Ghigna S., Governato F., Lake G.,
   %Quinn T., Stadel J., Tozzi P.,
   1999, ApJ, 524, L19
   %Dark Matter Substructure within Galactic Halos
   %related to the missing satellite problem
%We use numerical simulations to examine the substructure
%within galactic and cluster mass halos that form within
%a hierarchical universe. Clusters are easily reproduced
%with a steep mass spectrum of thousands of substructure
%clumps that closely matches the observations. However,
%the survival of dark matter substructure also occurs on
%galactic scales, leading to the remarkable result that
%galaxy halos appear as scaled versions of galaxy clusters.
%The model predicts that the virialized extent of the Milky
%Way's halo should contain about 500 satellites with circular
%velocities larger than the Draco and Ursa Minor systems,
%i.e., bound masses >~10^8 M_solar and tidally limited sizes
%>~1 kpc. The substructure clumps are on orbits that take a
%large fraction of them through the stellar disk, leading
%to significant resonant and impulsive heating. Their
%abundance and singular density profiles have important
%implications for the existence of old thin disks, cold
%stellar streams, gravitational lensing, and indirect/direct
%detection experiments.

\bibitem{Na} Navarro J., Steinmetz M., 1997, ApJ, 478, 13
%ionization feedback

\bibitem{} Niederste-Ostholt M., et al.
  %Belokurov V., Evans N. W., Gilmore G.,
  %Wyse R. F. G., Norris J. E.,
  2009, MNRAS, astro-ph/0906.3669

\bibitem{Pi} Piatek S., et al.
  %Pryor C., Olszewski E. W., Harris H. C., Mateo M.,
  %Minniti D., Monet D. G., Morrison H., Tinney C. G.,
2002, ApJ, 124, 3198
%determine proper motion of dSphs
%(dwarf spheroidals)

\bibitem{PU} Putman M. E., et al.
    %Grcevich J., Peek J. E. G.,
    2008, astro-ph/08033069
% HI content of Segue 1

\bibitem{Qu} Quinn T., Katz N., Efstathiou G., 1996, ApJ, 278, 49
%ionization feedback

\bibitem{Ri} Ricotti M., Gnedin N. Y., 2005, ApJ, 629, 259
%reionization scenario

\bibitem{Rie} Riehm T., et al.
  %Zackrisson E., Mortsell E., Wiik K.,
  2009, astro-ph/0905.4738
%millilensing dwarf galaxies

\bibitem{Si} Simon J. D., Geha M., 2007, ApJ, 670, 313
%solving missing satellite problem

\bibitem{So} Somerville R. S., 2002, ApJ, 572, L23
%reionization scenario

\bibitem{St} Strigari L. E., et al.
    %Bullock J. S., Kaplinghat M., Simon J. D.,
    %Geha M., Willman B., Walker M.,
    2008, Nature, 454, 1096
    %2. method of mass determination

\bibitem {Wa} Walker M. G., et al.
   %Mateo M., Olszewski E. W., Bernstein R., Wang X., Woodroofe M.,
   2006, AJ, 131, 2114
   % method to determined v_r^s, sigma^s

\bibitem{Wal} Walsh D., Carswell R. F.,
  Weymann R. J., 1979, Nature, 279, 381
%1. detection of gravitational lensing

\bibitem{We} Weinberg D. H., Hernquist L., Katz N., 1997, ApJ, 477, 8
%ionization feedback

\bibitem{Willman} Willman B., et al. 2005, ApJ, 626, L85
%In the 2000' there are also some observations finding a large
%number of satellites in the Local Group, mainly with the results
%from the Sloan Digital Sky Survey (SDSS; e.g., Willman et al.
%2005; Zucker et al. 2006; Belokurov et al. 2006). Perhaps, we
%could also take Willman et al. 2005 as an example in the 2000'.
%Their paper: Willman, B. et al. 2005, ApJ, 626, L85.

\bibitem{WuLou2006}Wu Y., Lou Y.-Q., 2006, MNRAS, 372, 992
%``Scale-Free Thin Discs in an Isopedic Magnetic Field", Wu Y., Lou
%Y.-Q., {\it Monthly Notices of the Royal Astronomical Society},
%{\bf 372}, 992-1018, 2006 (astro-ph/0607360).

\bibitem{Xi} Xiang-Gruess M., Lou Y.-Q.,
  Duschl W. J., 2009, MNRAS, 397, 815
%astro-ph/0905.2262
%``Global Non-axisymmetric Perturbation Configurations
% in a Composite Disc System with an Isopedic Magnetic Field: Relation
% between Dark Matter Halo and Magnetic Field", Xiang-Gruess M.,
% Lou Y.-Q., Duschl W. J., {\it Monthly Notices of the Royal
% Astronomical Society}, {\bf 397}, 815-936, 2009.

\bibitem{Za1} Zackrisson E., et al.
  %Riehm T., M\"oller O., Wiik K., Nurmi P.,
  2008, ApJ, 684, 804
  %millilensing dwarf galaxies

\bibitem{Za} Zackrisson E., Riehm T., 2009, astro-ph/0905.4075
%gravitational lensing dwarf galaxies

\bibitem{Zw1} Zwicky F., 1937a, Phys. Rev., 51, 290
% gravitational lensing, galaxies
%
\bibitem{Zw2} Zwicky F., 1937b, Phys. Rev., 51, 679
% gravitational lensing, galaxies

\end{thebibliography}
\end{document}